\documentclass[sigconf]{acmart}
\pdfoutput=1

\def\BibTeX{{\rm B\kern-.05em{\sc i\kern-.025em b}\kern-.08emT\kern-.1667em\lower.7ex\hbox{E}\kern-.125emX}}
    
\usepackage{soul,color}
\usepackage{adjustbox}
\usepackage{graphicx}

\usepackage{caption}
\usepackage{subcaption}
\usepackage{enumitem}

\graphicspath{{./pdf/}{./jpeg/}{./images/}}
 \DeclareGraphicsExtensions{.pdf,.jpeg,.jpg,.png}
 \DeclareGraphicsExtensions{.pdf,.jpeg,.jpg,.png}

\begin{document}




\title[Automated Classification of Incriminating Digital Forensic Artefacts]{Methodology for the Automated Metadata-Based Classification~of~Incriminating~Digital~Forensic~Artefacts}


%
\author{Xiaoyu Du}
\authornote{UCD Forensics and Security Research Group - \url{https://www.forensicsandsecurity.com}}
\email{xiaoyu.du@ucdconnect.ie}
\affiliation{%
  \institution{University College Dublin}
  \city{Dublin}
  \country{Ireland}
}

\author{Mark Scanlon}
\authornotemark[1]
\email{mark.scanlon@ucd.ie}
\affiliation{%
  \institution{University College Dublin}
  \city{Dublin}
  \country{Ireland}
}



 




%
\renewcommand{\shortauthors}{Du and Scanlon}

%
\begin{abstract}

The ever increasing volume of data in digital forensic investigation is one of the most discussed challenges in the field. Usually, most of the file artefacts on seized devices are not pertinent to the investigation. Manually retrieving suspicious files relevant to the investigation is akin to finding a needle in a haystack. In this paper, a methodology for the automatic prioritisation of suspicious file artefacts (i.e., file artefacts that are pertinent to the investigation) is proposed to reduce the manual analysis effort required. This methodology is designed to work in a human-in-the-loop fashion. In other words, it predicts/recommends that an artefact is likely to be suspicious rather than giving the final analysis result. A supervised machine learning approach is employed, which leverages the recorded results of previously processed cases. The process of features extraction, dataset generation, training and evaluation are presented in this paper. In addition, a toolkit for data extraction from disk images is outlined, which enables this method to be integrated with the conventional investigation process and work in an automated fashion.
\end{abstract}

%
%


%


\copyrightyear{2019} 
\acmYear{2019} 
\setcopyright{acmlicensed}
\acmConference[ARES '19]{Proceedings of the 14th International Conference on Availability, Reliability and Security (ARES 2019)}{August 26--29, 2019}{Canterbury, United Kingdom}
\acmBooktitle{Proceedings of the 14th International Conference on Availability, Reliability and Security (ARES '19), August 26--29, 2019, Canterbury, United Kingdom}
\acmPrice{15.00}
\acmDOI{10.1145/3339252.3340517}
\acmISBN{978-1-4503-7164-3/19/08}

\maketitle

\section{Introduction}
\label{sec: introduction}


``Big digital forensic data'' is a challenge faced by law enforcement agencies around the world as the prevalence of digital devices ever increases~\cite{lillis2016current}. The volume of data requiring analysis during digital forensic investigations is ever increasing. How to quickly detect pertinent file artefacts is a problem calling to be solved. A variety of solutions such as automation, data reduction, centralised digital evidence processing, data deduplication, triage, etc., have been proposed and developed in recent years.

Beebe et al.~\cite{beebe2009digital} outlined how the automation of forensic processes is important. Numerous automation tools have been developed by researchers in the field. However, independent automatic tools are showing very little impact on the efficiency of the investigative process. As the formats used by these tools vary, this leads to the issue that investigators are left to manually figure out the clues from the analyses results~\cite{case2008face}. Merging the disparate formatted results from various tools is also overly arduous and time consuming~\cite{chen2019security}. The Cyber-Investigation Analysis Standard Expression (CASE)\footnote{https://caseontology.org/}, a community-developed standard format, is attempting to bridge this gap.

For reducing the digital forensic backlogs, the first operational Digital Forensics as a Service (DFaaS) system, Hansken, was implemented by Netherlands Forensics Institute (NFI) in 2014~\cite{van2014digital, van2015digital}. DFaaS is a cloud based solution, analysing forensic data and sharing the results on a centralised sever. It enables easier information sharing between investigators and detectives, which is one manner that improves overall efficiency over traditional systems. Based on the DFaaS paradigm, data deduplication was proposed for reducing the time repeatedly acquiring and analysing known file artefacts~\cite{scanlon2016deduplication}. Several experiments have been performed that proves the increased efficiency created by a deduplication system~\cite{watkins2009teleporter, neuner2015gradually, du2018deduplicated}. Creation of centralised artefact whitelisting eliminates the known, benign operating system and application files, and also eliminates known, benign user created files. In addition, known illegal file artefacts can be detected at an earliest stage possible, i.e., during the acquisition phase.

Machine learning approaches offer a data driven solution, which results in a more flexible solution compared with hard coded scripts. In recent years, libraries such as \texttt{scikit-learn}~\footnote{https://scikit-learn.org/} facilitate simple and efficient data mining and data analysis. Researchers in many fields look to machine learning techniques to improve the effectiveness and efficiency of their solutions; the same is true in the digital forensics and cybersecurity domains. For example, machine learning has been employed in malware classification~\cite{le2018deep}, establishing forensic analysis priorities~\cite{grillo2009fast}, automated categorisation of digital media~\cite{marturana2013machine}, etc. 

The research presented in this paper aims to improve the automation of the digital forensic investigative process. To achieve this goal, a methodology is proposed to automatically determine suspicious/relevant file artefacts encountered during the investigation. Machine learning models were trained to perform this prediction task. A DFaaS system saved the expert human artefact classification, i.e., illegal or benign, on the server, which provides training data for the models. The experiments presented in this paper prove the viability of this methodology. 


\subsection{Contribution of this Work}

The contribution of this work can be summarised as:
\begin{itemize}[leftmargin=0.5cm]

\item Design of an automatic digital forensic data processing approach to prioritise the investigation of suspicious file artefacts;

\item Verification of the proposed methodology through an example scenario;

\item Creating a toolkit for data extraction form disk images, associated dataset generation,and pre-processing. This enables the method to be performed automatically during the investigation;

\item Analysis and discussion of the proposed solution in the field for accelerating the processing of large volumes of digital evidence.
    
\end{itemize}

\section{Related Work}
\label{relatedwork}
Section~\ref{sec: automation} to~\ref{sec: machine_learning} presents the existing methods and techniques on expediting the process of digital forensic investigation and combating the backlogs caused by the big data volumes. In addition, how these techniques can be combined during the investigation are discussed. In Section~\ref{sec: metadata_and_timeline}, the significant value of metadata and timeline analysis to digital forensic investigation is outlined, and associated tools and frameworks towards automatic metadata and timeline analysis are presented.

\subsection{Automatic Analysis}
\label{sec: automation}
Automating the investigative process is challenging. Garfinkel~\cite{garfinkel2010digital} pointed out that automation comes at great expense and has had limited success. The path followed in a digital forensic investigation can vary substantially due to the investigation's purpose, the type of case, and the variety of devices encountered. A number of process models have been defined, yet there is still no global standard procedure in practice. Hence, there is no existing, completely automated approach to conduct investigations.

Numerous automatic tools have been developed in the research and commercial fields that can assist the investigation process. These tools aim to only reduce part of manual work during evidence processing. Many automatic tools focus on data extraction. However, automatic artefact examination is more challenging.  

Attempting to chain existing automated tools achieves very limited progress to achieve this goal. This is due to the tools requiring specific formats for input data, and the format of generated output data can be incompatible with the next desired tool in the chain. Automatic processing is needed to improve the efficiency and reduce inadvertent errors in digital forensic investigation~\cite{braekt2016workflow}. In addition, as described in the introduction, the large volume of data leads to individual examiners being unable to fully understand it or use it effectively. One arduous task that investigators have to perform is connecting the dots~\cite{case2008face}.

Chen~\cite{chen2019security} outlines that merging analyses results from various tools into a single case report is difficult. Hence, the automation should be designed on a framework level, which means from the start to the end of the process, the tools work together automatically. 

Efforts have been made toward a higher level of automatic digital evidence processing. CASE aims to aggregate the extracted information by various forensic tools in a standardised manner. \texttt{log2timeline}\footnote{https://github.com/log2timeline/plaso/wiki} is a tool that generates a ``super timeline'', which consists of digital events from the various files and logs discovered. \texttt{log2timeline} uses parsers to extract digital events from a variety of file formats (e.g., NTFS \texttt{\$MFT}, Microsoft IIS log files, SQLite databases, Firefox Cache, Chrome preferences, etc.). However, existing analysis plug-ins provide very limited options.

\subsection{DFaaS and Data Deduplication}

The term \textit{Big Data Forensics} was proposed as a new branch of digital forensics by Zawoad et al.~\cite{zawoad2015digital}. In the age of Big Data, there are both new challenges and opportunities for digital forensics. Lillis et al.~\cite{lillis2016current} outlined the challenges faced by investigators; and one of the most impactful is the ever increasing data volumes requiring forensic analysis. Significant potential benefits for digital evidence processing can be seen from a number of techniques such as data mining, machine learning, and big data analytics.

In 2014, the Hansken DFaaS system was described by the NFI~\cite{van2014digital}. The system provides a service that processes high volumes of digital material in a forensic context. This system had processed over a petabyte of data, as of 2015~\cite{van2015digital}. As the benefits outlined in ~\cite{van2015digital}, DFaaS improves the efficiency of the investigation through offering better resource management, collaboration and sharing knowledge, etc.

One question to be noted is how DFaaS can work with the existing forensic process models. The DFaaS system from the NFI is based on the \textit{Integrated Digital Forensic Process Model}~\cite{kohn2013integrated}. In 2017, Du et al.~\cite{du2017evaluation} discussed the evolution of digital forensic process models, as well as an overview of the benefits of DFaaS to existing process models. DFaaS is no longer a new paradigm and is compatible with traditional evidence processing frameworks.

Employing data deduplication techniques to combat the big data volume and expedite digital evidence processing has been discussed in the literature~\cite{watkins2009teleporter, scanlon2016deduplication, zawoad2015digital}. For applying deduplication techniques, a centralised database recording the known file artefacts is required. This technique can work on top of the DFaaS framework. The employment of data deduplication techniques effectively reduces the data required to be processed during the investigation, and potentially blacklisting enables a faster illegal file artefact detection~\cite{scanlon2016deduplication, neuner2015gradually}. Experimentation has proven the significant savings of the storage and the volumes of data processed in~\cite{neuner2015gradually} and~\cite{du2018deduplicated}.

\subsection{Prioritised Analysis}
\label{sec:Prioritised}
Even with the large number of devices and data involved in modern investigations, the forensic value of each artefact is not equal. Conducting comprehensive analysis on all seized digital devices is overly time consuming. How to quickly determine which data has more pertinent value for manual examination is an open research question in existing literature.

Triage is a term originally used in medical field; in digital forensics, triage ranks seized digital devices in terms of importance or priority~\cite{rogers2006computer}. Digital forensic triage is a method usually used for getting faster responses to an incident for time-sensitive cases, such as child abuse, kidnapping, and terrorist threats~\cite{hitchcock2016tiered}. When the volume of seized material is very large and only few devices might be considered relevant for the investigation, how to quickly determine which are the devices with the higher forensic value is paramount.

Prioritised analysis on file artefacts has become necessary in a variety of cases. As the storage of each device is increasing, the number of file artefacts on a device could be huge. Manual examination of each file artefact could take a long time. Conventional methods rely on the hash-based filtering and keyword indexing to search for relevant file artefacts. In recent years, researchers have proposed metadata-based clustering methods for grouping similar file artefacts to assist the investigators. Document clustering has also been used for forensic analysis~\cite{da2013document}.

\subsection{Machine Learning in Digital Forensics}
\label{sec: machine_learning}
Machine learning-based techniques have been widely applied across diverse fields. Deep learning, as another subcategory of artificial intelligence, really shines when dealing with complex problems such as image classification, nature language processing, and speech recognition. A number of methods have been proposed to develop intelligent methods for problem solving in digital forensics. 

An approach for computer user identification was presented by Grillo et al. in order to quickly classify seized devices~\cite{grillo2009fast}. The proposed method for identifying the individual computer user leveraged the user's habits, computer skill level, online interests, etc. User profiling resulted in five categories being identified: occasional users, Internet chat users, office worker users, experienced users, and hacker users. This approach prioritised the analysis of seized hard drives; forensic examiners could examine only potentially relevant hard drive images resulting in a reduced analysis time.

Triage in forensic investigation has been discussed in Section~\ref{sec:Prioritised} and using machine learning technique to conduct triage on seized devices was discussed. Marturana et al.~\cite{marturana2013machine} presented a triage method for the categorisation of digital media. The features extracted from the seized devices were based on the connections to specific crimes under investigation. Two use cases were presented in this paper; one involved copyright infringement, while the other involved child sexual abuse investigation. The crime-related features were defined by the researchers, which resulted in models being relevant for the case categories defined, but may not be applicable to real cases directly.

\subsection{Metadata Forensics and Timeline Examination}
\label{sec: metadata_and_timeline}

Typical questions often asked in digital forensic investigation include: when, where, what, why, and how the incident happened~\cite{casey2011digital}. Metadata refers to data about data and it usually plays significant role in digital forensic investigation. The precise information contained in metadata varies depending on the file system and file type. Generally, metadata consists of file modification time stamps, file ownership information, and data units allocated to this metadata unit~\cite{altheide2011digital}. In digital forensic investigation, file system metadata is typically used for 1) keyword searching to find out a specific type of file artefact, e.g., based on the file type; 2) filtering the file artefacts based on file size, creation time, and so on~\cite{garfinkel2010digital}. 

In recent years, metadata has also been applied for automatic forensic investigation. Rowe and Garfinkel~\cite{10.1007/978-3-642-35515-8_10} outlined a methodology that uses directory metadata (file names, extensions, paths, size, access/modification times, fragmentation, status flags and hash codes) on a large corpus to find out anomalous and suspicious file artefacts. In 2013, Raghavan and Raghavan~\cite{raghavan2013determining} demonstrated the use of metadata associations to determine the origin of downloaded files.

Digital forensic analysis attempts to reconstruct the events that have occurred on the seized device(s) pertaining to a case. Creating a timeline of system activity is capable of assisting the investigators in the discovery of useful traces for the case. Inglot et al. claimed that there is a need for a comprehensive timeline analysis tool~\cite{inglot2012framework}.

Hargreaves et al.~\cite{hargreaves2012automated} proposed an automated timeline reconstruction approach for generating high-level events, which reduced the number of events to be analysed. One problem highlighted by these authors is that separate neural networks are required to be trained for different applications and a new data set of file system activity would be required for newer versions of applications. More general purpose analysers are needed for digital event correlation.  

\texttt{log2timeline and plaso}~\footnote{https://github.com/log2timeline/plaso/wiki}~\cite{gudhjonsson2010mastering} appears to be the tool of choice for timeline generation and analysis over recent years. It is a tool designed to extract timestamps (of digital events) from various of log files found on a typical computer system and aggregate them. 
\texttt{log2timeline} analysers conduct the preliminary discovery of the digital events contained on a device. However, analysers generating useful information through digital events correlation must be developed to enable an automated evidence analysis process.

\section{Methodology}
\label{methodology}

\begin{figure*}[!ht]
\centering
\includegraphics[width=\textwidth]{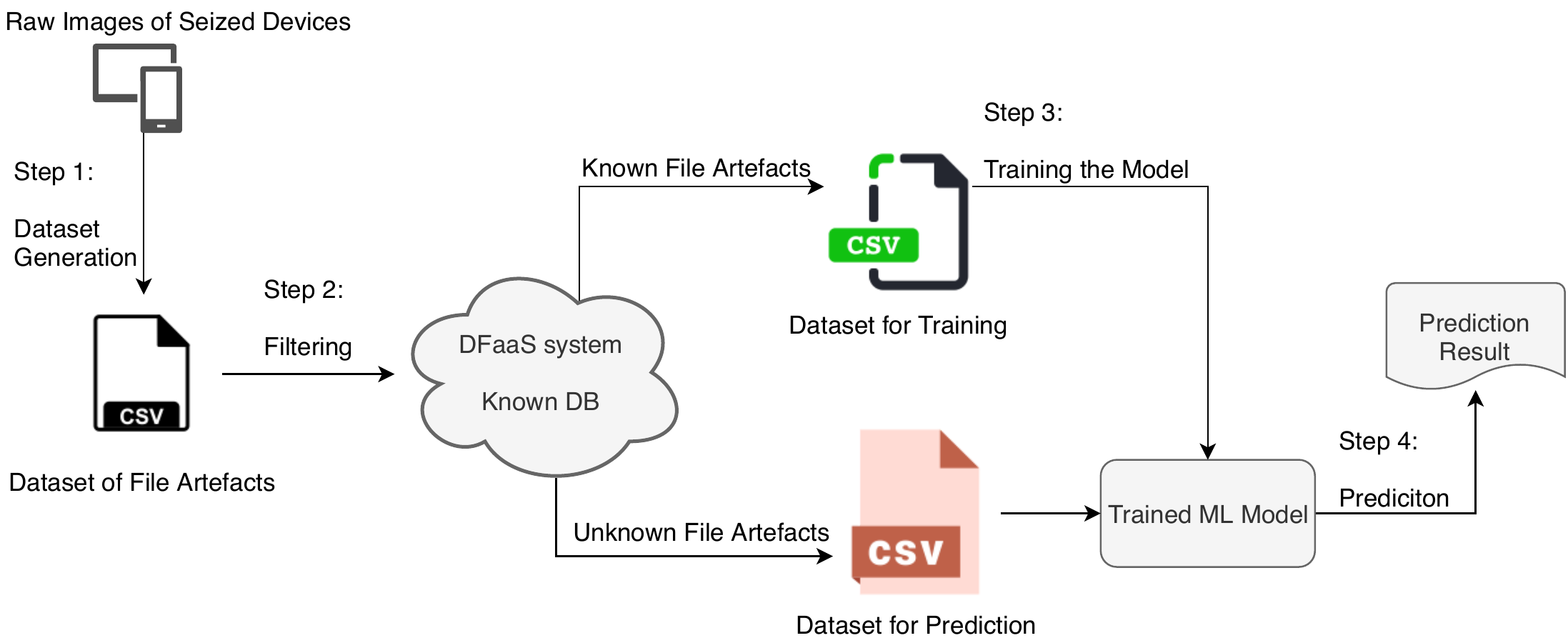}
\caption{Overview of Methodology: the work-flow of the proposed methodology in an investigation.}
\label{fig: overview}
\end{figure*}
Traditional, hash-based white-listing or blacklisting methods are usually the go-to automatic solution for finding known file artefacts during a digital forensic investigation. If nothing pertinent has been detected, the investigators will have to manually perform keyword search or filter to examine the artefacts. Using hash matching to detect illegal files can only detect the precisely same file artefacts or artefacts with a minor change (i.e., by using approximate matching). 

This research presents an approach that enables automatic determination of suspicious file artefacts by training machine learning models. The associated metadata and digital events occurred are employed to extracting features of each file artefact. Combining this with a centralised, deduplicated digital evidence processing framework, illegal file artefacts encountered in previous cases are labelled as such in the database. These files on the blacklist can be used to train classifiers for detecting previous unencountered suspicious file artefacts in new cases.



\subsection{The Relevancy Determination Process}
\label{sec: relevancy}

Figure~\ref{fig: overview} illustrates the designed work-flow of this methodology as could be applied in an investigation. From a raw format image copy to the prediction result, the process is completely automated. The processing is broken down in four steps:

\begin{enumerate}

\item \textbf{Dataset Generation} - The developed toolkit is used to generate the training dataset in \textit{csv} format for machine learning-based modelling. The details on how this dataset is generated is presented in Section~\ref{sec: toolkit}.

\item \textbf{Filtering} - It works through comparing the hash with the known database, which will split the file artefacts into two categories; known file artefacts and unknown file artefacts.

\item \textbf{Training the Model} - Training a model using known file artefacts. The features used will be described in Section~\ref{sec: features}.

\item \textbf{Prediction} - The categorisation of the previously unencountered files are predicted by the trained model (i.e., if it is relevant or suspicious to the investigation).

\end{enumerate}

Through the above process, from the raw image input, an initial automatic analysis result can be performed automatically. The output is a prediction of each artefact as suspicious or not to assist the investigator in identifying the file artefacts likely relevant to the investigation.

\subsection{Toolkit for Data Extraction and Processing}
\label{sec: toolkit}

\begin{figure}[!ht]
\centering
\includegraphics[width=0.5\textwidth]{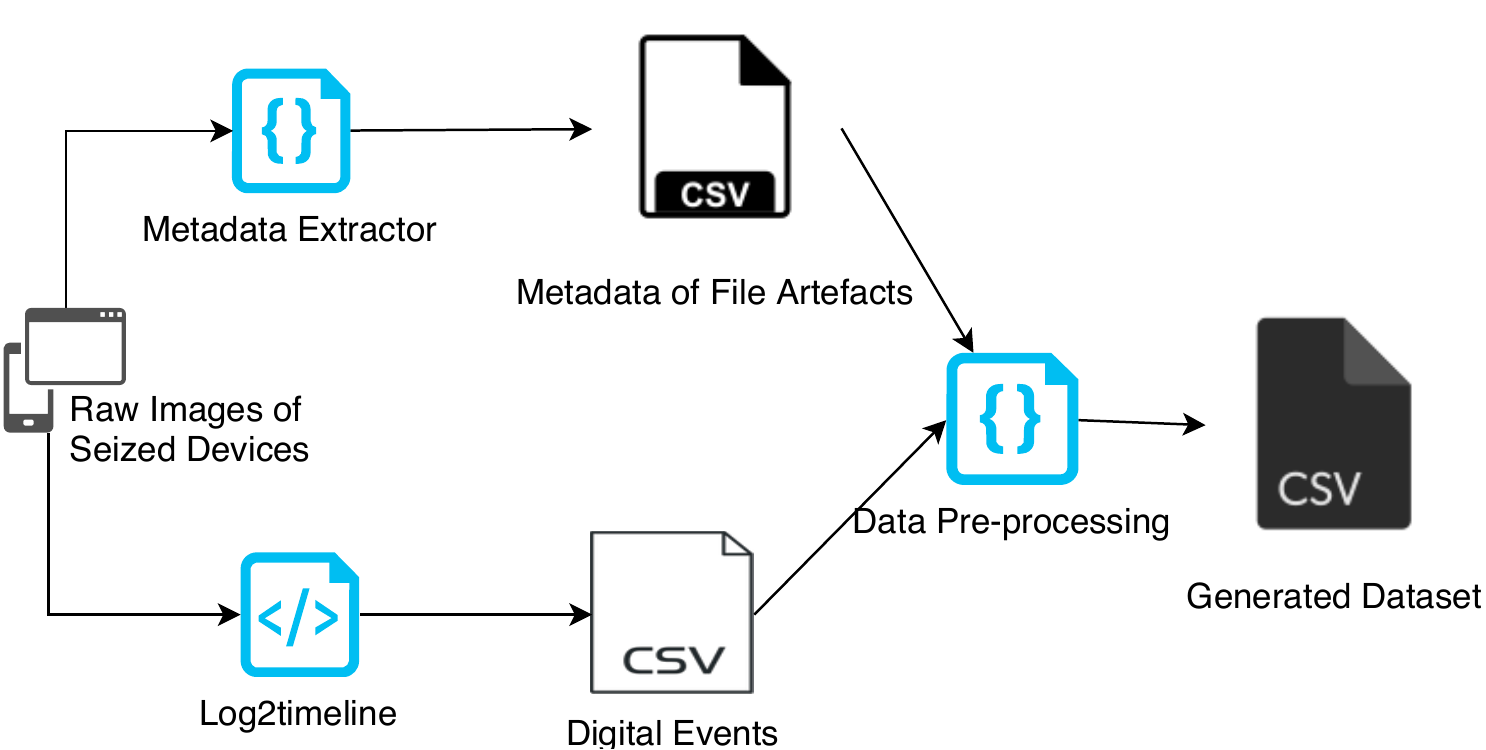}
\caption{Toolkit for Data Extraction and Processing}
\label{fig:toolkit}
\end{figure}

Figure~\ref{fig:toolkit} shows how the dataset for training is generated from a raw disk image. There are three parts at this stage: 1) metadata extraction using the python library \texttt{pytsk}~\footnote{https://github.com/py4n6/pytsk}; 2) digital events extraction using \texttt{log2timeline}; 3) merging the metadata and digital events to a dataset for file artefacts. The tools for metadata and digital events extraction operate directly on raw disk image. The input is disk image in \textit{raw} format. The output is a file in \textit{csv} format.

The metadata extraction tool uses \texttt{pytsk}. \texttt{Pytsk} is a Python binding for the Sleuth Kit. In this research, it was used to extract the file system metadata. The timeline generated using \texttt{log2timeline} consists of the digital events from the file system, windows registry, browsing history, download history, etc. Each event describes source, the file name, event type, etc. 

In the generated timeline, each row represents a digital event and the~\texttt{inode} can identify which file artefact it is related to. The~\texttt{inode} can be used to collate all of the digital events associated with each file artefact. If multiple partitions are involved, a combination of hash and~\texttt{inode} must be used to keep this unique identifier. The combination of metadata and timeline results in an abundance of information related to each file artefacts. More details on the extracted data is presented in Section~\ref{sec:metadata_and_timeline}.  

\subsection{File System Metadata and Timeline}
\label{sec:metadata_and_timeline}

The importance of metadata analysis in an investigation process has been discussed in Section~\ref{sec: metadata_and_timeline}. The specific file system metadata considered useful and exploited in this research are:
\begin{itemize}[leftmargin=0.5cm]
    \item File name;
    \item File size;
    \item \texttt{inode};
    \item File hash value (MD5/SHA1/SHA256);
    \item Physical address of file;
    \item The number of blocks of the file;
    \item The block number(s) allocated to the file;
    \item Creation time of the file;
    \item Last access time of the file;
    \item Last modification time of the file.
\end{itemize}

Timelines contain more information than just timestamps. The ``super timeline'' generated by \texttt{log2timeline} consists of what happened, when it happened, on which artefact, and where each digital event was recorded on the system. A complete list of the fields of generated timeline are categorised and listed as follows:

\begin{itemize}[leftmargin=0.5cm]

\item Fields describing the event:

\begin{itemize}[leftmargin=0.5cm]
\item \textbf{date} - Date that the event occurred;
\item \textbf{time} - Time that the event occurred;
\item \textbf{MACB} - Modification, access, creation and birth times;
\item \textbf{desc} - A description of the timestamp object;
\item \textbf{short} - A shorter version description of the timestamp;
\item \textbf{filename} - The file object on which the event occurred;
\item \textbf{sourcetype} - Description of the source type, e.g., ``Opera Browser History'';
\item \textbf{source} - A shorted form of the source, e.g., ``WEBHIST'';
\item \textbf{type} -Timestamp type, e.g., ``Last Time Executed''.
\end{itemize}

\item Fields describing the source:
\begin{itemize}[leftmargin=0.5cm]
\item \textbf{inode} - The inode or MFT number of the parsed artefacts;
\item \textbf{user} - The user owns the parsed artefacts;
\item \textbf{host} - The host that the data came from.
\end{itemize}

\item Fields describing the tool used:
\begin{itemize}[leftmargin=0.5cm]
\item \textbf{version} - the version of the tool;
\item \textbf{timezone} - Timezone where applying the tool generating the timeline;
\item \textbf{format} - The parsing module.
\end{itemize}

\item Fields describing other information:
\begin{itemize}[leftmargin=0.5cm]
\item \textbf{notes} - An operational field;
\item \textbf{extra} - A reference to a hash that stores all additional fields that might be used.
\end{itemize}
\end{itemize}

\subsection{Feature Extraction from Metadata and Timeline}
\label{sec: features}

In training machine learning models, not every feature available proves valuable. For example, randomly generated numerical features, artefact hash values, or \texttt{inode} values are not helpful to the prediction task. Feature manipulation is usually needed for a specific task or purpose for each machine learning model. Trough feature transformation, the information input for model training can be added, changed or removed as desired for each task. In fact, the resultant model is a way of constructing a new feature that solves the task at hand~\cite{flach2012machine}. 

Standalone metadata, such as filename and timestamp, are likely not suitable to use directly for training classification models. Table~\ref{table: feature} lists the useful metadata and the corresponding features that can be extracted, manipulated, and transformed. 

\begin{table}[h!]
\caption{Valuable Feature Extracted}
\centering
\begin{tabular}{ |p{2cm}||p{5cm}|}
\hline
\textbf{Metadata} & \textbf{Features Extracted} \\
\hline\hline
Timestamp &  Day of the week, time of day, the number of years/months/days/hours of file created/last access/modified, etc. \\
\hline
 Filename & Length of filename, character types in the filename, language, etc.  \\
\hline
File Type  & Type of file, for example, image, document, executable, etc. This is based on file header information as opposed to file extensions.  \\
\hline
Owner & Username of the file creator.  \\
\hline
File Size & Categorising the size in KB. \\
\hline
File Path & Depth of the directory; depth is defined as the number of parent directories in the hierarchical filesystem.\\
\hline
Digital Events & The number of associated events occurred on the file artefacts in total; the number of the events occurred on the file artefacts on/in a specific day/week/month; the most frequent type/source of events happened; and so on.
\\

\hline

\end{tabular}

\label{table: feature}

\end{table}


\subsection{Evaluation Matrix}
\label{sec:evaluation}

Due to the severe imbalance of the dataset, accuracy is not used to evaluate the performance. Because a model can predict the value of the majority class for all predictions and achieve a high classification accuracy, this model is not useful in the problem domain.

The performance metrics used are precision, recall and F1-score:

\begin{itemize}
\item \textbf{Precision} is the fraction of relevant instances among the retrieved instances: 
    
\begin{center}
\textit{precision = TP/(TP + FP)};  
\end{center}
    
\item \textbf{Recall} is the fraction of relevant instances that have been retrieved:
\begin{center}
\textit{recall = TP/(TP + FN)};
\end{center}

\item \textbf{F1-Score} is an overall measure of a model's accuracy that combines precision and recall:
\begin{center}
\textit{F1-Score = 2*recall*precision}.  
\end{center}
\end{itemize}

\subsection{Comparison with the Existing Methodology}
\label{sec: comparing}
Differentiating this approach from previous research on using metadata to cluster the file artefacts, this research leverages expert human analysis results from the manual processing previous investigations. The hypothesis that the features of analysed file artefacts can enable trained machine learning models to determine how relevant newly encountered file artefacts are to a specific type of investigation. The experiment presented in Section~\ref{experiment} explores this hypothesis. 

One advantage of the proposed method is performing supervised machine learning tasks on the investigated file artefacts; the automated categorisation can more directly steer the investigator's focus towards pertinent data at the earliest possible stage. Lastly, this research presents a complete overview of how this method can be applied in real-world investigative scenario.

\section{Experimentation and Results}
\label{experiment}

This section outlines the dataset generation, the structure of the dataset, the machine learning models trained and the performance of the results are evaluated.


\begin{figure}[!ht]
\centering
\includegraphics[width=0.43\textwidth,trim={1cm 0 1cm 0}]{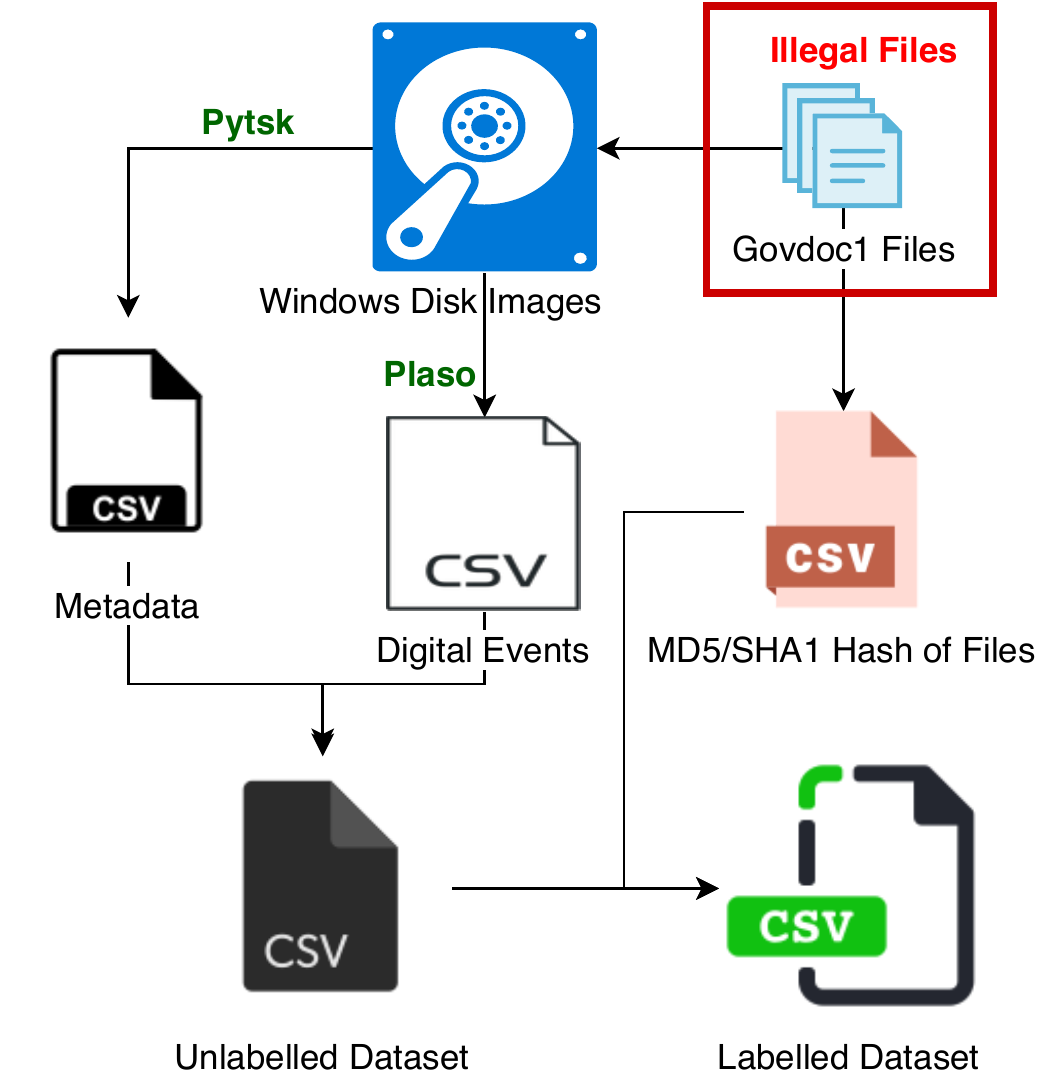}
\caption{Dataset Generation - 1) Disk image creation; 2) Metadata and timeline generation; 3) Merging the two sources (each sample in the dataset represents a single file artefact); 4) Labelling the data (file artefacts from \texttt{Govdoc1} Corpus are defined as ``illegal'').}
\label{fig:dataset_method}
\end{figure}

\subsection{Experimentation Dataset}

In this section, the creation of dataset used in the experiment is outlined. As shown in figure~\ref{fig:dataset_method}, the process of experimental dataset generation includes: 
\begin{enumerate}[leftmargin=0.5cm]
    \item Emulating wear-and-tear of the device on a virtual environment and planting the ``illegal'' file artefacts on the test virtual machine;
    \item  Using the developed tool to extract filesystem metadata and a ``super timeline'' from the disk image;
    \item  Merging the information about file artefacts from two sources (i.e., the extracted timeline and metadata);
    \item  Labelling the file artefacts on the dataset based on the hash of file artefacts.
    
\end{enumerate}

The disk images used in this research are Windows 7 disk images. Emulated wear and tear actions (surfing the Internet, installing software, downloading files from browsers, etc.) were conducted in a virtual environment. The \texttt{Govdocs1}\footnote{https://digitalcorpora.org/corpora/files} dataset were defined as ``known illegal'' file artefacts (\texttt{Govdocs1} consists of almost 1 million freely-redistributable files of various formats and sizes).

\subsection{Example Scenario}

In child abuse material possession/distribution investigation cases, pertinent digital evidence often consists of multimedia content with similar file size, under same directory, very close creation time, last access time, etc. The proposed methodology aims to detect suspicious files in such investigative scenarios. Hence, file artefacts with a similar number of associated events, file type, similar path on the device with known illegal file artefacts should likely be predicted as suspicious/relevant.

Based on the purpose of the prediction task, the complete matrix of the features are:

\begin{itemize}[leftmargin=0.5cm]
\item \textbf{Depth of Dir} - Integer representing the depth of the file directory (i.e., the number of parent directories);
\item \textbf{File Extension} - Categorical data type;
\item \textbf{Length of Name} - An integer representing the filename's length.
\item \textbf{Creation Time (y)} - How many years old is the file;
\item \textbf{Creation Time (m)} - How many months old is the file;
\item \textbf{Creation Time (d)} - How many days old is the file;
\item \textbf{Creation Time (h)} - How many hours old is the file;
\item \textbf{Size} - The data size of file in KB;
\item \textbf{Count} - The number of associated file events;
\item \textbf{Class} - If the file benign or illegal. (the value is 0 for the benign files, 1 for the illegal files).
\end{itemize}

\begin{figure*}[!ht]
\centering
\includegraphics[width=\textwidth]{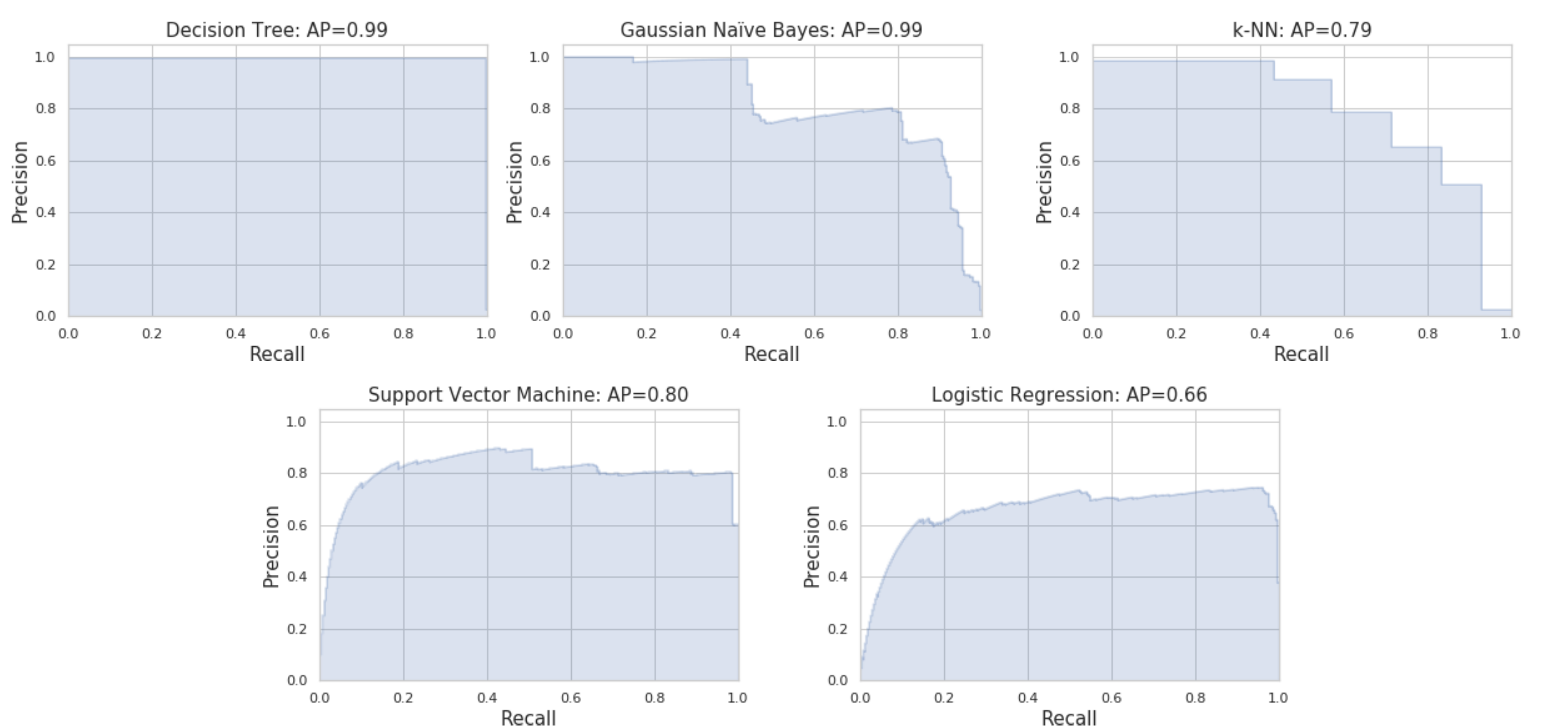}
\caption{Precision-Recall Curves per Classifier with Corresponding Average Precision (AP) Scores}
\label{fig:RP}
\end{figure*}

\subsection{Result Discussion}

Several machine learning classification algorithms including Logistic Regression, k-NN, Support Vector Machine, Decision Tree, Gaussian Na\"ive Bayes are tested in the experimentation. The datasets used in the experiment were generated using the toolkit presented in Section~\ref{sec: toolkit}. Table~\ref{table: dataset} shows the generated datasets used in this experiment; mainly differing on the percentage of illegal artefacts. The dataset is split into training and testing data. 

\begin{table}[!ht]
\begin{center}
\begin{tabular}{|c |c| c| c|}
\hline
 Dataset &  No. of Artefacts & Benign Artefacts &Illegal Artefacts  \\
\hline \hline
Dataset1 & 42,326 & 41,339 &987 \\
\hline
Dataset2 & 55,296 &49,328 & 5,968 \\
\hline
\end{tabular}
\end{center}
\caption{Datasets Used in the Experiment}
\label{table: dataset}
\end{table}

Precision-Recall curves summarise the trade-off between the true positive rate and the positive predictive value for a predictive model using different probability thresholds. For getting an overall understanding of the models' performance, Precision-Recall curve (RP curve) is visualised and average precision score (AP score) is calculated. Figure~\ref{fig:RP} shows the RP curve of the employed algorithms on \texttt{dataset1}. Comparison of the average precision score, the performance of models from best to worst are Decision Tree, Gaussian Na\"ive Bayes, Support Vector Machine, k-NN, and Logistic Regression respectively.




As the practical usage of the trained model is for digital forensic investigation, more concern should be put on the classifying of illegal files (class 1). Precisely, the precision, recall and F1 score on class 1, instead of the average value on class 0 and class 1, should be used to represents the performance of the models. The scores shown in Table~\ref{table: result_matrix} are on the class 1. This table shows an evaluation matrix from two datasets.

As shown in Table~\ref{table: result_matrix}, it is obvious that the performance with \texttt{dataset2} is significantly better than \texttt{dataset1}. This happens is because the number of ``illegal'' file samples in \texttt{dataset2} is more than \texttt{dataset1}. The imbalanced class problem is apparent in these datasets. The datasets were created in this manner due to the assumption that only a subset of the ``illegal'' files are classified as known illegal/known benign from the centralised database.


\begin{table*}[!ht]
\begin{center}
\begin{adjustbox}{width=1\textwidth}
\small
\begin{tabular}{c||c|c|c||c|c|c||c}
 \hline
Algorithms &\multicolumn{3}{c||}{datsaset1} &\multicolumn{3}{c||}{datsaset2} & Short Summary \\

    & precision & recall & f1-score & precision & recall & f1-score & \\ 
    \hline
    Decision Trees & 0.99 & 1.00 & 0.99 & 1.00 & 1.00 &1.00 & Best models in this experiment \\
    
Gaussian Na\"ive Bayes & 0.16 & 0.97 & 0.27 & 0.99 & 1.00 & 0.99 & Performance influenced by the dataset very much\\ 

 k-NN & 0.79 & 0.71 & 0.75 & 1.00 & 1.00 & 1.00 & Better performance on \texttt{dataset2}\\
  Support Vector Machines & 0.82 & 0.52 & 0.64 & 1.00 & 1.00 &1.00 & Better performance on \texttt{dataset2} \\
  
  Logistic Regression & 0.71 & 0.67 & 0.69 & 0.99 & 1.00 & 0.99 & Better performance on \texttt{dataset2}\\ 

  \hline
\end{tabular}
\end{adjustbox}
\end{center}
\caption{Evaluation Matrix of Different Classification Algorithms and Datasets}

\label{table: result_matrix}
\end{table*}

An overview on the models' settings and performance analyses, as shown in Figure~\ref{fig:RP}, is listed below:

\begin{itemize}[leftmargin=0.5cm]

\item \textbf{Decision Tree} - Observing from the AP score and precision, recall and F1 scores, the Decision Tree classifier performs best for both datasets compared against other models. In addition, the performance is stable on the different datasets.

\item \textbf{Gaussian Na\"ive Bayes} - The performance of Gaussian Na\"ive Bayes on \texttt{dataset1} is worse than \texttt{dataset2}. Even though the AP score is good, the scores on class 1 is much worse. For this severely imbalanced dataset, the score for the class with less samples is very low.

\item \textbf{k-NN} - One important parameter for k-NN is the selection of \texttt{k}. In this experiment, the model achieved its best accuracy when \texttt{k} was set to be 5.

\item \textbf{Support Vector Machine} - Through experimentation, the model realises a poor performance when the given data is not normalised. Hence, the dataset is normalised before it is fed into the model whose parameters are left default for the experiment.

\item \textbf{Logistic Regression} requires quite large sample sizes, which could be the reason it scores lower comparing with the other models.
\end{itemize}

The performance of these models indicates that the proposed methodology is valid and justifies further exploration. Among the aforementioned algorithms, Decision Tree achieved the best performance. The percentage distribution of the file artefacts among the different classes can be various in the real-world investigation. Different distribution of illegal file and benign file dataset should be tested for giving further conclusion which model is more suitable for specific scenarios.

\section{Conclusion}
\label{conclusion}

This paper outlines an automatic approach to use machine learning models to predict which file artefacts are likely pertinent to an investigation (i.e. which file artefacts are likely more suspicious than others). It is designed to integrate with a DFaaS framework, rather than a stand alone experiment on an individual device. The associated toolkit was introduced that was developed for supporting the automation of some the digital forensic process. An example scenario was outlined and tested indicating the feasibility and effectiveness of the proposed methodology. The promising experimental results for suspicious artefact detection provides motivation for further research.

\subsection{Future Work}

\label{futurework}
Future work in this ares will focus on:

\begin{itemize}[leftmargin=0.5cm]

\item Generating more scenarios - This experimental scenario used in this paper was child abuse material possession and distribution investigation. The purpose of this case type is to find out files with similar size, directory, creation time, etc. More investigative scenarios will be designed and evaluated. For example, training models for determining suspicious files through each origin source (from email attachment, Cloud account, USB device, etc.), third party owner, etc.  

\item More exploration is required on the extent of imbalanced classes influences each model's performance. Form the conducted experiment in this paper, it was shown that the dataset can influence the performance of the models. In the future work, more diverse datasets will be generated for testing. The generation of these datasets should provide a broad variety on the ratios of the benign/illegal files.

\item Exploring appropriate feature selection and feature engineering approaches for the defined machine learning tasks. In this paper, the features are selected by the common knowledge of digital forensic investigation. One avenue of exploration that could improve the performance is extracting as many features as possible initially, and subsequently selecting from them by using techniques such as \textit{SelectBest}, \textit{RFE (Recursive Feature Elimination)}, and \textit{PCA (Principal Component Analysis)}.

\item Tuning the model so that it can fit needs of real investigative scenarios better. For example, to focus on the recall scores, rather than insisting on a higher precision. Because, in any digital forensic investigation, any pertinent inculpatory or exculpatory file artefact can not be inadvertently overlooked.

\end{itemize}




%
\bibliographystyle{ACM-Reference-Format}
\bibliography{sample-base}

%
\end{document}